\documentclass[10pt,twoside,BCOR7mm,DIV15,headinclude,footexclude,
               cleardoubleempty,idxtotoc]
{scrartcl}

\usepackage{natbib}
\usepackage[font=small,labelfont=bf]{caption}
\usepackage[english]{babel}
\usepackage{graphicx}
\usepackage{hyperref}
\usepackage{scrpage2}
\usepackage{ifthen}
\usepackage{booktabs}
\usepackage{amsmath}
\usepackage{amssymb}
\usepackage{multicol}
\usepackage{float}
\usepackage{hyperref}

\hypersetup{breaklinks=true
,colorlinks=true,linkcolor=black,urlcolor=blue
,citecolor=black}

\addto\captionsenglish{%
}

\makeatletter
\renewcommand{\@biblabel}[1]{}
\renewcommand{\@cite}[2]{%
{#1\ifthenelse{\boolean{@tempswa}}{,#2}{}}}
\makeatother
\ofoot{\thepage}
\ifoot{}

\setcapindent{0em}
\setheadsepline{1pt}

\setkomafont{pagehead}{\normalfont\sffamily}
\setkomafont{pagenumber}{\normalfont\rmfamily}

\makeatletter
\newcommand{\listofcontributions}{\@starttoc{con}}

\newcommand{\l@contribution} {\@dottedtocline{1}{1.5em}{2.3em}}
\makeatother

\newenvironment{contribution}{
\setcounter{section}{0}
\setcounter{figure}{0}
\setcounter{table}{0}
}{
\newpage
\lehead{}
\rohead{}
}

\def\apj{ApJ}%
\def\apjl{ApJ}%
\def\apss{Ap\&SS}%
\def\aap{A\&A}%
\def\aapr{A\&A~Rev.}%
\def\aaps{A\&AS}%
\def\nat{Nature}%
\begin{document}

\setlength{\baselineskip}{2.5ex}

\begin{contribution}

\lehead{Jorick S. Vink}

\rohead{The True origin of Wolf-Rayet stars}

\begin{center}
{\LARGE \bf The True origin of Wolf-Rayet stars}\\
\medskip

{\it\bf Jorick S. Vink$^1$}\\

{\it $^1$Armagh Observatory, College Hill, BT61 9DG Armagh, Northern Ireland, UK}\\

\begin{abstract}
The Wolf-Rayet (WR) phenomenon is widespread in astronomy. It involves 
classical WRs, very massive stars (VMS), 
WR central stars of planetary nebula CSPN [WRs], and supernovae (SNe). 
But what is the root cause for a certain type of object to turn into an emission-line star? In this contribution, I discuss 
the basic aspects of radiation-driven winds that might reveal the ultimate difference 
between WR stars and canonical O-type stars. 
I discuss the aspects of (i) self-enrichment via CNO elements, 
(ii) high effective temperatures ($T_{\rm eff}$), (iii) an increase in the helium abundance ($Y$), and finally (iv) 
the Eddington factor $\Gamma_{\rm e}$. Over the last couple of years, we have made a breakthrough in our 
understanding of $\Gamma_{\rm e}$-dependent mass loss, which will have far-reaching consequences for the 
evolution and fate of the most massive stars in the Universe.
Finally, I discuss the prospects for studies of the WR phenomenon in the highest 
redshift Ly$\alpha$ and He {\sc ii} emitting galaxies.
\end{abstract}
\end{center}

\begin{multicols}{2}

\section{Introduction}

\begin{figure*}
\begin{center}
\includegraphics
  [width=\textwidth]{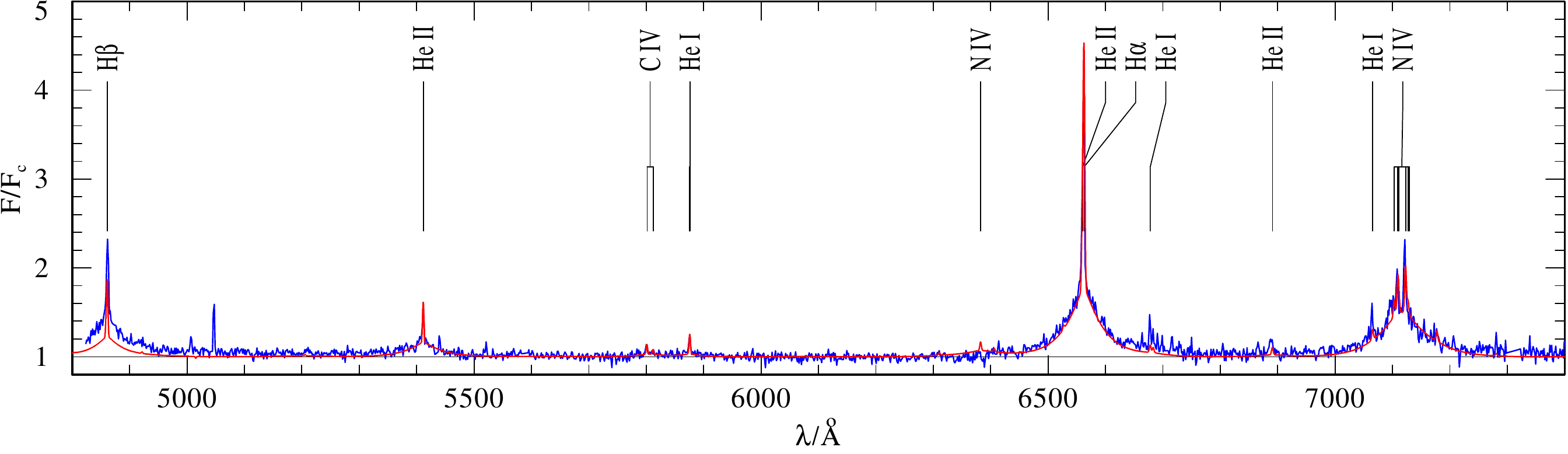}
\caption{Model fit to the observed spectrum of the IIb supernova SN 2013cu, 15.5 hours after 
explosion (Gal-Yam et al. 2014).
See Gr\"afener \& Vink (2015b) for more details on the model fit.}
\end{center}
\end{figure*}

Before being able to address the question regarding the true origin of 
Wolf-Rayet (WR) stars, we need to ask ourselves what exactly do we mean with the 
term ``origin'', as it could potentially refer to an evolutionary origin, as 
classical WR stars are thought to be evolved massive stars. 
It is thus important 
to emphasize that evolutionary status is not the only factor that 
decides whether a given object is expected to give rise to the WR phenomenon.
Instead, what I am interested in here is which 
physical mechanism is responsible for the WR phenomenon?
In other words, what is the root cause for a certain type of 
object to reveal the strong emission lines caused by recombination cascades? 

There are basically {\it four} types of objects which we would wish to explain 
by the same mechanism. These four groups are: (i) the 
evolved WR stars of type WN/WC/WO (nitrogen, carbon, oxygen-rich respectively), 
(ii) WNh stars, which are very 
massive stars (VMS) with $M > 100 M_{\odot}$ (Vink 2015a), (iii) WR central stars
of planetary nebula CSPN [WRs], and since very recently also (iv) young supernovae (SNe) such 
as SN\,2013cu, which reveal WR-like spectra, as depicted in Fig. 1.

Because of the WR phenomenology in the latter case of SN 2013cu,  
Gal-Yam et al. (2014) proposed a WR progenitor system, however, whilst 
the Type IIb SN 2013cu indeed shows a WR-like emission-line spectrum, 
the narrow lines are more consistent with the slow winds of Luminous 
Blue Variables (LBVs), as has been discussed by Groh (2014) and Gr\"afener \& 
Vink (2015b). This is also more consistent with the fact that LBVs were first 
suggested to be the variable-wind progenitors of Type IIb SNe with the additional 
evidence of their much slower winds than WR stars (Kotak \& Vink 2006). 

In this contribution, I discuss several properties of WR stars that may 
potentially make them distinct from the more common O-type stars. 
These involve (i) self-enrichment with large amounts of
carbon (in WC stars) and oxygen (in WO stars); (ii) high
effective temperatures ($T_{\rm eff}$) ; (iii) an enhanced helium ($Y$) abundance, and finally, 
(iv) a high luminosity-to-mass ratio ($L/M$), as represented by the Eddington 
factor $\Gamma_{\rm e}$:  

\begin{equation}
\Gamma_{\rm e}~=~g_{\rm rad}^{\rm elec}/g_{\rm grav}~=~\sigma_{\rm elec} L/(4 \pi c G M)
\end{equation}
where $g_{\rm rad}^{\rm elec}$ is the radiative acceleration on electrons, and the 
other symbols have their usual meaning.
I will be assuming some basic knowledge of radiation-driven 
wind theory, but if the reader wishes to learn more details, 
I refer to Puls et al. (2008), Owocki (2015) and Vink (2015b).

\section{Self-enrichment (CNO)}

\begin{figure*}
\begin{center}
\includegraphics[width=\textwidth]{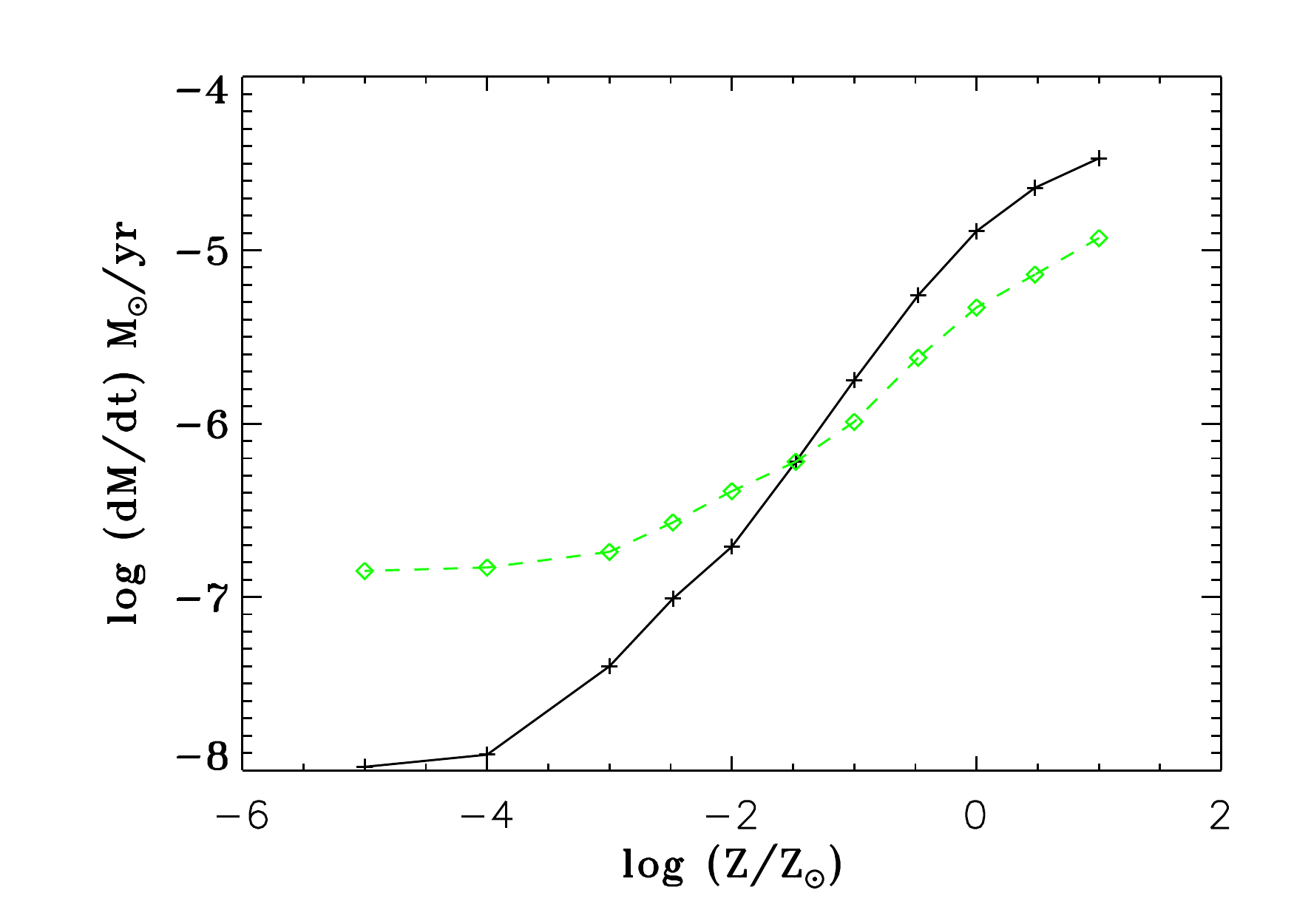}
\caption{Monte Carlo WR mass-loss predictions as a function of metallicity $Z$. The 
dark line represents the late-type WN stars, whilst the lighter dashed line shows 
the results for late-type WC stars. The slope for the WN models is similar to the 
predictions for OB-supergiants, whilst the slope is
somewhat shallower for WC stars. At lower Z, the slope becomes smaller, whilst it flattens off entirely at
extremely low $Z$. These computations are from Vink and de Koter (2005).}
\end{center}
\end{figure*}

Classical WR stars are evolved objects that have lost a significant amount of 
their initial mass (50 - 90\,\%) due to stellar winds or 
alternative mass-loss (either single star or binary) during earlier evolutionary phases. 
WC/WO stars are significantly enriched in C and O (e.g. Sander et al. 2012; Tramper et al. 2015; 
Georgy et al. 2012; Yoon et al. 2012).

For decades it was assumed that the high metal ($Z$) content in classical WC/WO atmospheres 
was responsible for the high mass-loss rates due to $Z$-dependent stellar winds, where
$Z$ includes CNO elements. 
For instance, the oft-used empirical mass-loss formula for classical WR stars by Nugis \& Lamers (2000)
includes a $Z$-dependence, where $Z$ primarily refers to the abundant CNO elements. 
It was not until 2005, that it became clear that 
classical WR stars do in fact show an iron (Fe) $Z$-dependence (see Fig.\,2) -- where $Z$ is now the Fe abundance of the host galaxy -- 
due to the millions of Fe lines that drive the inner winds (Vink \& de Koter 2005; Gr\"afener \& Hamann 2008).

This strong dependence on Fe, and implicit weak dependence on CNO elements in the inner wind, is very similar 
in O-type stars (Vink et al. 1999; Puls et al. 2000), which implies that 
self-enrichment cannot be the prime reason for the WR phenomenon in the local 
Universe (this might be different at very low $Z$ where CNO line-driving takes over; Vink \& de Koter 2005).

\begin{figure*}
\begin{center}
\includegraphics[width=\columnwidth]{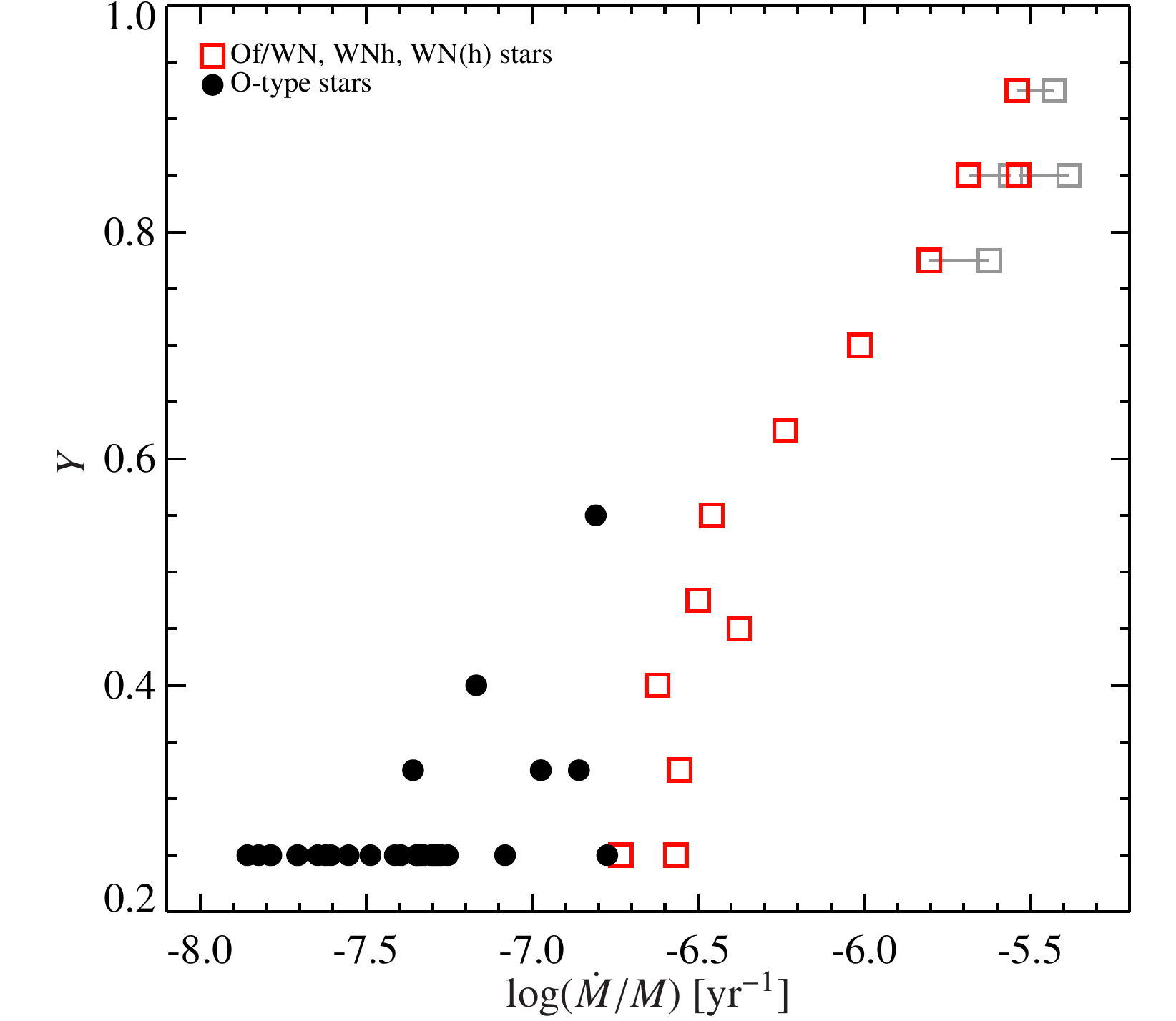}
\caption{He surface mass fractions $Y$ vs.\ relative
  mass-loss rates $\log(\dot{M}/M)$ for the VMS in VFTS by Bestenlehner et al. (2014)
  (normal O-type stars are indicated with black circles). 
  Masses have been estimated using the assumption
  of chemical homogeneity (from Gr\"afener et al. 2011). I.e. they are upper limits. 
  Potential
  core He-burning masses are also indicated (in light grey).}
\end{center}
\end{figure*}

\section{High temperatures $T_{\rm eff}$}

Classical WR stars are very hot with high core temperatures and high wind densities.
These properties result in a well-stratified ionization structure (Lucy \& Abbott 1993), with the 
special properties that gaps in one atmospheric/wind layer become effectively ``blocked''
by the emergence of a new set of lines from a lower ionization stage. This physical
mechanism involving photon trapping (e.g. Owocki 1994) leads to a very efficient momentum transfer via 
highly efficient multiple scattering and high mass-loss rates. It is thus very tempting
to argue that the high temperatures are the main reason for the high mass-loss rates in 
WR stars. This argument becomes even stronger when one realizes that also [WR] stars
of central stars of planetary nebulae (CSPNe) are very hot, with mass-loss rates higher 
than O-type CSPNe (see Todt, these proceedings).

However, the more recent discovery and modelling of WNh VMS stars in young clusters like the Arches cluster and 
S Dor (Martins et al. 2008; Bestenlehner et al. 2014) 
has revealed that their temperatures are not in any way special with respect to those of 
more common lower-luminosity O-type stars. 
The only real difference is that they are more luminous, and thus more massive, which
explains why they are oftentimes
referred to as ``O-stars on steroids''. 

The fact that these moderately-hot WR stars (of order 50\,000 K instead of 
100\,000 - 200\,000 K for classical WRs) can still entertain 
strong stellar outflows above the single-scattering limit (Vink \& Gr\"afener 2012) indicates 
that high temperatures alone cannot be the sole reason for the WR phenomenon either.

\section{High helium abundance ($Y$)}

Due to its atomic structure, helium has more spectral lines available than hydrogen.
This simple fact could thus potentially play a role 
in providing more efficient wind driving. 
Furthermore, the underlying spectral energy distribution (SED)
of a helium star is different from that of a main-sequence hydrogen burning 
object, which could potentially also  
affect the efficiency of radiative driving. 
Furthermore, the empirical mass-loss formula of Nugis \& Lamers (2000) 
shows a clear dependence on $Y$. Finally, recent results in the context of the 
VLT Flames Tarantula survey (VFTS) of the most massive stars in 30 Dor by Bestenlehner et al. (2014) 
reveal higher mass-loss rates for higher $Y$ abundance (see Fig. 3).

However, I would find it highly unlikely that high $Y$ content is the key to 
the WR phenomenon.
Already in 2002, Monte Carlo predictions for LBVs showed that the 
differences in radiation driven wind mass loss due to $Y$ enrichment were only 
subtle (Vink \& de Koter 2002). 
It is therefore far more likely that there is no causal relationship between 
$Y$ and $dM/dt$ as Fig. 2 might naively suggest, but that helium enrichment 
is actually the {\it result} of a high-mass loss rate 
instead (Bestenlehner et al. 2014). 

\begin{figure*}
\begin{center}
\includegraphics[width=\textwidth]{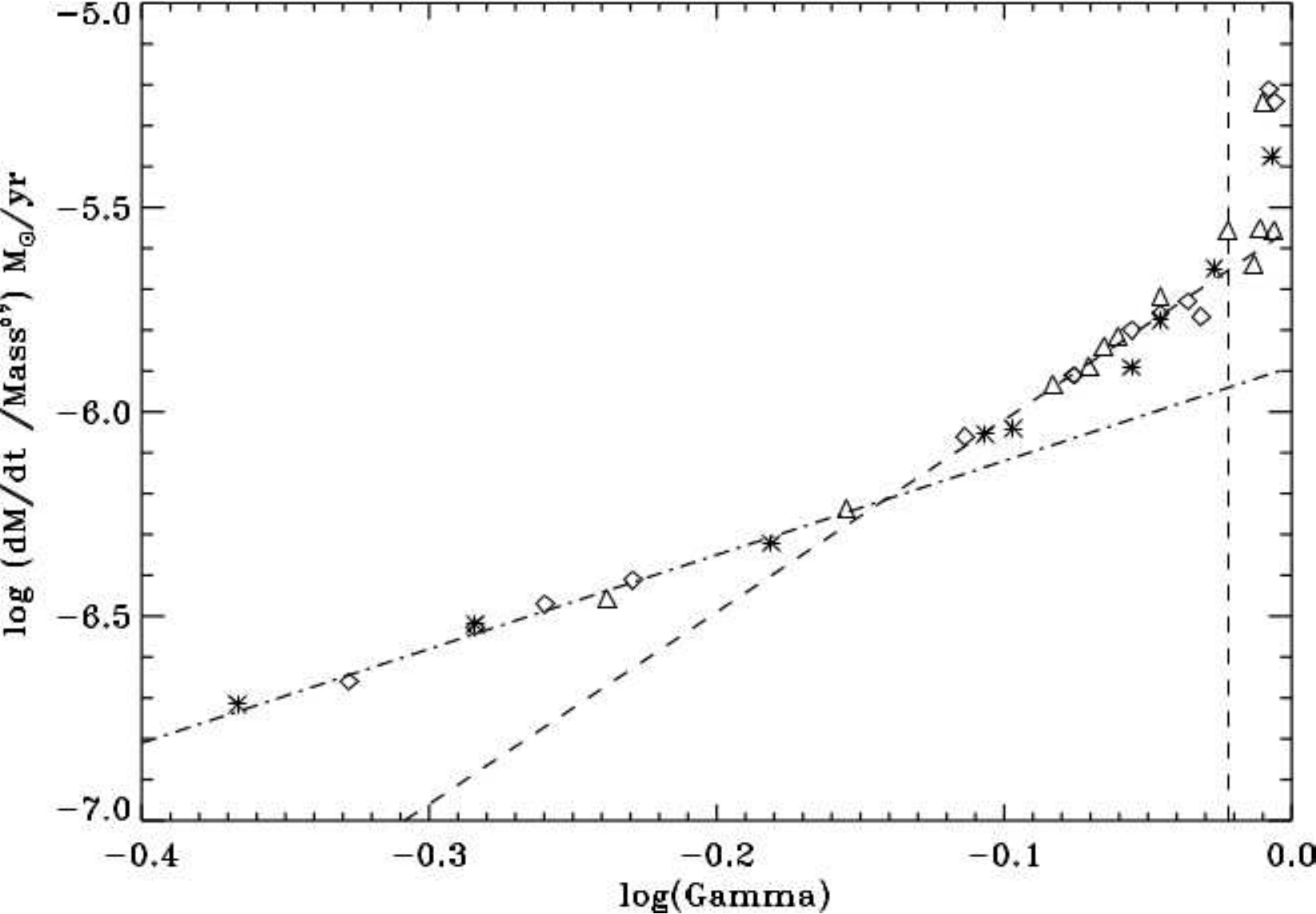}
\caption{Monte Carlo mass-loss predictions (divided by$M^{0.7}$) 
versus $\Gamma_{\rm e}$ for models approaching the Eddington limit. 
The dashed-dotted line represents the best linear fit for the 
range $0.4 < \Gamma_{\rm e} < 0.7$. 
The dashed line represents the higher $0.7 <  \Gamma_{\rm e} < 0.95$ 
range (from Vink et al. 2011). The {\it kink} occurs where the 
winds become optically thick, and where the single-scattering 
limit is crossed simultaneously.}
\end{center}
\end{figure*}

\section{Proximity to Eddington limit} 

The results in Fig. 3 may easily be explained by the fact that the kink in Fig. 3 
is seen at exactly the same position as the kink in Fig. 4.
Figure 4 shows Monte Carlo mass-loss predictions for VMS including both optically thin
O-type stars as well as optically thick WNh stars. 

The slope in the relationship switches 
abruptly from 2 to 5 when the optical depth crosses unity (Vink \& Gr\"afener 2012). 
It is noteworthy to emphasize that whilst the shallow slope of the canonical O-type stars 
is consistent with classical Castor, Abbott \& Klein (1995) (hereafter CAK) radiation-driving, the steep
slope is clearly inconsistent with classical CAK-type radiation driven winds for optically thin winds.
Instead, these optically thick winds are determined by physics beyond CAK (see Bestenlehner et al. 2014 for 
a more extensive discussion as to whether the kink could alternatively also be reproduced by 
(modified) CAK theory with variable force multipliers). 

I would like to note that the steep slope shown in Fig.\,4 is also found in independent 
calculations by Gr\"afener \& Hamann (2008), as well as empirical results 
by Gr\"afener et al. (2011) and Bestenlehner et al. (2014) for VMS in the Arches cluster and 30 Dor, respectively. 
Given these latest results on VMS stars it seems most likely that it is {\it the Eddington factor $\Gamma_{\rm e}$ 
that is the key to the WR phenomenon}, but only the future will be able to tell if this assertion is correct.

\section{Outlook on He {\sc ii} emission}

Helium {\sc ii} emission is now routinely detected in 
the highest redshifts Ly$\alpha$ emitting 
star-forming galaxies. 
VMS with WR phenomena likely play a pivotal role in this He {\sc ii} emission, as these hot
VMS are now known to dominate the ionizing radiation in local starbursts such as 30 Dor (Doran et al. 2013). 
Understanding the WR phenomenon in VMS is thus pivotal for understanding high-redshift galaxies, as well 
as the reionization of the Universe (Vink 2015a). 

It may not only be the ionizing radiation of VMS that is relevant for
Helium {\sc ii} emission: Cassata et al. (2013) recently published 
a study of He {\sc ii} emitters between redshifts 2 and 5 including some 
{\it narrow} He {\sc ii} emitters.
Note that broad He {\sc ii} emission is generally attributed to stellar emission from 
classical WR stars, whilst narrow He {\sc ii} emission is normally
attributed to nebular emission -- excited by hot stars.

However, Gr\"afener \& Vink (2015a) recently suggested an alternative scenario 
for the origin of narrow He {\sc ii} emission, namely the {\it slow} winds
of VMS at low $Z$, as a result of their proximity to the Eddington limit. 
This narrow He {\sc ii} emission from very early generations of VMS may
become detectable in studies of star-forming galaxies at high
redshifts with the James Webb Space Telescope (JWST). 

The WR phenomenon may thus still hold a number of future 
intriguing surprises!

\bibliographystyle{aa} 
\bibliography{myarticle}

\end{multicols}

\end{contribution}


\end{document}